\documentclass[aps,prl,twocolumn,showpacs,groupedaddress]{revtex4}  
\usepackage{graphicx}  
\usepackage{dcolumn}   
\usepackage{bm}        
\usepackage{amssymb}   

\usepackage{xspace}

\newcommand{\dzero}     {D0\xspace}
\newcommand{\ttbar}     {\ensuremath{t\overline{t}}\xspace}

\newcommand{\pythia}    {\sc{pythia}}

\newcommand{\alpgen}    {\sc{alpgen}}

\newcommand{\comphep}   {\sc{comphep}}
\newcommand{\met}       {\mbox{$\not\!\!E_T$}\xspace}

\newcommand{\wplus}     {$W$+jets\xspace}
\newcommand{\zplus}     {$Z+$jets\xspace}

\newcommand{\muplus}    {$\mu$+jets\xspace}
\newcommand{\eplus}     {$e+$jets\xspace}
\newcommand{\lplus}     {$\ell+$jets\xspace}

\newcommand{\sigmaB}    {$\sigma_{X}\!\cdot\! B(X\!\rightarrow\!\ttbar)$}
\newcommand{\TeV}{\,\mbox{Te\kern-0.2exV}\xspace}
\newcommand{\GeV}{\,\mbox{Ge\kern-0.2exV}\xspace}
\newcommand{\ipb}{\,\mbox{pb}^{-1}}
\newcommand{\ifb}{\,\mbox{fb}^{-1}}
\newcommand{\pb}{\,\mbox{pb}}

\begin{document}

\hspace{5.2in} \mbox{Fermilab-Pub-08/097-E}

\title{\boldmath Search for $t\bar t$ resonances in the lepton plus jets final state\\ 
in $p\bar p$ collisions at $\sqrt{s}=1.96\TeV$}
%
\author{V.M.~Abazov$^{36}$}
\author{B.~Abbott$^{75}$}
\author{M.~Abolins$^{65}$}
\author{B.S.~Acharya$^{29}$}
\author{M.~Adams$^{51}$}
\author{T.~Adams$^{49}$}
\author{E.~Aguilo$^{6}$}
\author{S.H.~Ahn$^{31}$}
\author{M.~Ahsan$^{59}$}
\author{G.D.~Alexeev$^{36}$}
% ...

\affiliation{\vspace{0.1 in}(The D\O\ Collaboration)\vspace{0.1 in}}
% ...
\affiliation{$^{6}$University of Alberta, Edmonton, Alberta, Canada,
                Simon Fraser University, Burnaby, British Columbia, Canada,
                York University, Toronto, Ontario, Canada, and
                McGill University, Montreal, Quebec, Canada}
% ...
\affiliation{$^{29}$Tata Institute of Fundamental Research, Mumbai, India}
% ...
\affiliation{$^{31}$Korea Detector Laboratory, Korea University, Seoul, Korea}
% ...
\affiliation{$^{36}$Joint Institute for Nuclear Research, Dubna, Russia}
% ...
\affiliation{$^{49}$Florida State University, Tallahassee, Florida 32306, USA}
% ...
\affiliation{$^{51}$University of Illinois at Chicago,
                Chicago, Illinois 60607, USA}
% ...
\affiliation{$^{59}$Kansas State University, Manhattan, Kansas 66506, USA}
% ...
\affiliation{$^{65}$Michigan State University,
                East Lansing, Michigan 48824, USA}
% ...
\affiliation{$^{75}$University of Oklahoma, Norman, Oklahoma 73019, USA}
% ...
\date{April 23, 2008}

\begin{abstract}
We present a search for a narrow-width heavy resonance decaying into top
quark pairs ($X\rightarrow t\bar t$) in $p\bar p$ collisions at $\sqrt{s} =
1.96\TeV$ using approximately $0.9\ifb$ of data collected with the D0 detector at the
Fermilab Tevatron Collider.
This analysis considers \ttbar candidate events in the lepton plus jets 
channel with at least  one identified $b$ jet and uses  
the \ttbar invariant mass distribution 
to search for evidence of resonant production. 
We find no evidence for a narrow resonance $X$ decaying to \ttbar.
Therefore, we set upper limits on {\sigmaB} 
for different hypothesized resonance masses using a Bayesian approach.
For a Topcolor-assisted technicolor model, the existence of a leptophobic $Z'$ boson with 
mass $M_{Z'} < 700$\GeV\ and width $\Gamma_{Z'} = 0.012 M_{Z'}$ can be
excluded at the $95\%$ C.L.
\end{abstract}

\maketitle 

\section{\label{sec:intro}Introduction}
The top quark has by far the largest mass of all the known fermions. 
Unknown heavy resonances may play a role in the production of top quark
pairs  ($t\bar t$) and add a resonant part to the standard model (SM) production
mechanism mediated by the strong interaction.
Such resonant production is possible for massive $Z$-like bosons in
extended gauge theories \cite{Leike:1998wr}, Kaluza-Klein states of the gluon
or $Z$ boson~\cite{Lillie:2007yh,Rizzo:1999en}, axigluons \cite{Sehgal:1987wi}, Topcolor \cite{topc1}, and
other  theories beyond the SM.  
Independent of the exact model, resonant production of top quark pairs could
be visible in the reconstructed $t\bar t$ invariant mass distribution.

In this Letter, we present
a search for a narrow-width heavy resonance $X$ decaying into 
{\ttbar}.
We consider the lepton+jets ($\ell$+jets, where $\ell = e $~or~$\mu$) final
state.
 The event
signature is one isolated electron or muon with high  momentum transverse to
the beam axis ($p_T$), 
large transverse energy imbalance (\met) due to the undetected neutrino, and at least four jets, 
two of which result from the hadronization of $b$ quarks. 
The analyzed dataset corresponds to an integrated luminosity of
$913\pm 56\ipb$ in the \eplus 
channel and $871 \pm 53\ipb$ in the \muplus channel, collected with the D0
detector between August
2002 and December 2005. The
analysis uses events with  at least three reconstructed jets.
Backgrounds from light-quarks are further reduced by identifying $b$ jets.
After $b$ tagging, the dominant physics background for a resonance signal is 
non-resonant SM {\ttbar} production. Smaller contributions arise from
the direct production of $W$ bosons in 
association with jets (\wplus), as well as instrumental background originating from
multijet processes with jets faking isolated leptons. 
The search for resonant production in the \ttbar invariant mass distribution
is performed using Bayesian statistics to compare SM and
resonant production to the observed mass distribution.

Previous searches performed
by the CDF
and D0 collaborations  in Run~I
\nocite{RunIttrescdf} 
found no evidence for a \ttbar resonance~\cite{RunIttrescdf,RunIttresdzero}. 
In these studies, a Topcolor model was used as a reference to quote mass limits.
According to this model~\cite{topc1}, a large top quark mass can be
generated through the formation of a dynamical {\ttbar} condensate, $Z'$, due
to  a new strong gauge force with large coupling to the third
generation of fermions. In one particular model, Topcolor-assisted
technicolor~\cite{topc2}, the $Z'$ boson  
has large couplings only to the first and third generation of quarks
and has no significant couplings to leptons.
Limits obtained on
{\sigmaB} are used to set a lower bound on the mass of such a leptophobic
$Z'$ boson. In Run~I CDF found $M_{Z'}>480$\GeV with $106\ipb$ of data~\cite{RunIttrescdf}, 
and  {\dzero} obtained $M_{Z'}>560$\GeV using
$130\ipb$~\cite{RunIttresdzero}, both at the $95\%$ C.L. and for a 
resonance with width $\Gamma_{Z'} = 0.012 M_{Z'}$.

\section{\label{sec:detector}D0 Detector}
The D0 detector~\cite{Abazov:2005pn} has a central-tracking system consisting of a 
silicon microstrip tracker 
and a central fiber tracker, 
both located within a 2\,T superconducting solenoidal 
magnet, with designs optimized for tracking and 
vertexing at pseudorapidities $|\eta|<3$ and $|\eta|<2.5$,
respectively.
The pseudorapidity, $\eta$, is defined with respect to the beam axis.
{Central and forward preshower detectors are positioned just outside of 
the superconducting coil.} A liquid-argon and uranium calorimeter has a 
central section (CC) covering pseudorapidities $|\eta|$  
$\lesssim 1.1$, and two end calorimeters (EC) that extend coverage 
to $|\eta|\approx 4.2$, with all three housed in separate 
cryostats~\cite{Abachi:1993em}. An outer muon system covering $|\eta|<2$
consists of a layer of tracking detectors and scintillation trigger 
counters in front of 1.8\,T iron toroids, followed by two similar layers 
after the toroids~\cite{Abazov:2005uk}. 
{Luminosity is measured using plastic scintillator 
arrays placed in front of the EC cryostats. The three-level trigger and data 
acquisition systems are designed to accommodate the high luminosities  
of Run~II and record events of interest at up to
about $100$\,Hz.} 

\section{\label{presel}Event Selection}

To select top quark pair candidates in the \eplus and \muplus decay
channels,  triggers that required a jet and an electron or
muon are used.
The event selection requires either an isolated electron with  $p_T>20$\GeV\ and
$|\eta|<1.1$, or an isolated muon with $p_T>20$\GeV\ and $|\eta|<2.0$.
No additional  isolated leptons with $p_T>15$\GeV\ are allowed in the event.
Details of the lepton identification and isolation criteria are described
in~\cite{Abazov:2005ex,Abazov:2005ey}. 
We require {\met}
to exceed $20$\GeV\ ($25\GeV$) for the \eplus (\muplus) channel.
Jets are defined using a cone algorithm~\cite{Baur:2000bi} with radius 
${\cal R}_\mathrm{cone}=0.5$, where ${\cal R}_\mathrm{cone}=\sqrt{(\Delta \phi)^2 +(\Delta
 y)^2 }$, $\phi$ is the azimuthal angle, and $y$ the rapidity. 
The selected events must contain three or more jets
with $p_T>20$\GeV\ and $|y|<2.5$. At least one of the jets is
required to have  $p_T>40$\GeV.
Events with mismeasured lepton momentum
are rejected by requiring the $\met$ to be acollinear with the lepton direction
in the transverse plane:
$\Delta\phi(e,\met)>2.2-0.045\GeV^{-1}\,\met$ and $\Delta\phi(\mu,\met)>2.1-0.035\GeV^{-1}\,\met$~\cite{Abazov:2007kg}.

To improve the signal-to-background ratio, at least one jet is required
to be identified as a $b$ jet. The tagging algorithm uses the
impact parameters of tracks matched to a given jet  and
information on vertex mass, the decay length significance, and the number
of participating tracks  for any reconstructed secondary vertex
within the cone of the given jet. The information is combined in a  neural
network to obtain the output variable,
$\mathrm{NN}_B$, which tends towards one for
$b$ jets and towards zero for light quark jets~\cite{Scanlon:2006wc}.  In this
analysis we consider jets to be $b$-tagged if $\mathrm{NN}_B>0.65$
which corresponds to a tagging efficiency for $b$ jets of about 55\% with a
tagging rate for light quark jets of less than 1\%.

We independently analyze 
events with three and four or more jets and separate
singly tagged
and doubly tagged events, since
the channels have different  signal-to-background ratios and systematic
uncertainties.

\section{\label{bkgd}\label{mc}Signal and Background Modeling}

Simulated events are used to determine selection efficiencies for the 
resonant \ttbar production signal and for background sources except those
in which instrumental effects give fake leptons and \met in multijet
production events.
Samples of resonant \ttbar production are generated with 
{\pythia}~\cite{pythia63}
for ten different choices of the resonance mass $M_X$
between $350\GeV$
and $1$\TeV. In all cases, the width of the resonance is set to 
$\Gamma_X = 0.012 M_X$. This  qualifies the $X$ boson as a narrow resonance since its width is 
smaller than the estimated mass resolution
of the {\dzero} detector of 5--10\%. 
The generated resonance is forced to decay into \ttbar.

Standard model \ttbar and diboson backgrounds ($WW$, $WZ$, and $ZZ$)
are generated with {\pythia}~\cite{pythia63}.
Single top quark
production is generated
using the {\comphep} generator~\cite{Boos:2006af}.
A top quark mass of $175\GeV$
is used for both resonant and SM top production processes.
\wplus and \zplus events are generated using {\alpgen}~\cite{alpgen} to model the
hard interaction and {\pythia} for  parton showering, hadronization 
and hadron decays. 
To avoid double counting between the hard matrix element and the
parton shower, the MLM jet-matching algorithm is used \cite{MLM}. 
The  {CTEQ6L1} parton distribution
functions~\cite{Pumplin:2002vw,Stump:2003yu}  are used for all samples.
The generated events are processed through the full {\sc geant3}-based~\cite{GEANT} 
simulation of the D0 detector and the same reconstruction program as used for data.

The SM \ttbar, single top quark, diboson, and \zplus backgrounds are
estimated completely from Monte Carlo (MC) simulation, 
to obtain the total acceptance as well as the shape of the
reconstructed 
\ttbar invariant mass distribution. 
Trigger inefficiencies and differences between data and MC lepton and jet identification
efficiencies are accounted for by
weighting the simulated events~\cite{Abazov:2007kg}.  Jet $b$-tagging probabilities are
measured in data and parametrized as
functions of $p_T$ and $\eta$. They are used  to weight each simulated event
according to its event $b$-tagging probability.
Finally, the expected yields are
normalized to the SM theoretical prediction. 
A \ttbar production of
$\sigma_{\ttbar}=6.77 \pm 0.60\pb$ for $m_t=175\GeV$~\cite{topcs} is
used. \zplus,  single top quark and diboson samples are normalized to their next-to-leading-order 
cross sections~\cite{Hamberg:1990np,Sullivan:2004ie,Campbell:1999ah}.

The $W$+jets background is estimated from a combination of data and
MC information. The expected number of $W$+jets events in the  $b$-tagged sample is computed as
the product of the estimated number of $W$+jets before  $b$ tagging and the expected
event  $b$-tagging probability. 
The former is obtained from the observed number
of events with real leptons in data, computed using the matrix method~\cite{Abazov:2005ex},
and then subtracting the expected contribution from other SM
production processes. 
The  $b$-tagging probability is obtained by combining the \wplus flavor fractions 
estimated from MC with the event  $b$-tagging 
probability, estimated from $b$ tag rate functions.
The shape of the reconstructed invariant mass distribution is obtained from the MC simulation.

The multijet background is completely determined from data. The total number of expected events
is estimated by applying the matrix method to the each of the $b$-tagged subsamples. 
The shape is derived from events with leptons
failing the isolation requirements.
A summary of the prediction for the different background contributions in the combined $\ell$+jets 
channels, along with the observed number of events in data, is given in
Table~\ref{tab:yields-ljet-tagged}. Systematic uncertainties are discussed  below.

\begin{table}[bt]
\caption{\label{tab:yields-ljet-tagged}Event yields for the expected SM
  background and for data. The uncertainties are statistical and systematic.
  }
\begin{tabular}{lr@{}llr@{}ll}
\hline                      
\hline                      
                           & \multicolumn{3}{l}{~~~~~$3$ jets~} & \multicolumn{2}{l}{~$\ge4 $ jets~}\\
\hline                      
$t\bar t$                  & 167&.4   &&  160&.5&\\
\wplus                     &  118&.2  &&  24&.1&\\
Other MC                   &  34&.8   && 9&.8&\\
Multijet                   & 31&.3    && 7&.4&\\
\hline
Total background           & 351&.7&$\!\!\pm$\,29.3&  201&.8&$\!\!\pm$\,29.0\\
\hline
Data                       &~~~~ 370&&&  ~~~~237&\\
\hline
\hline                      
\end{tabular}
\end{table}

\section{\label{invmass}\boldmath Reconstruction of the $t\bar{t}$ invariant mass distribution}
\begin{figure*}[t]
\includegraphics[width=0.495\textwidth]{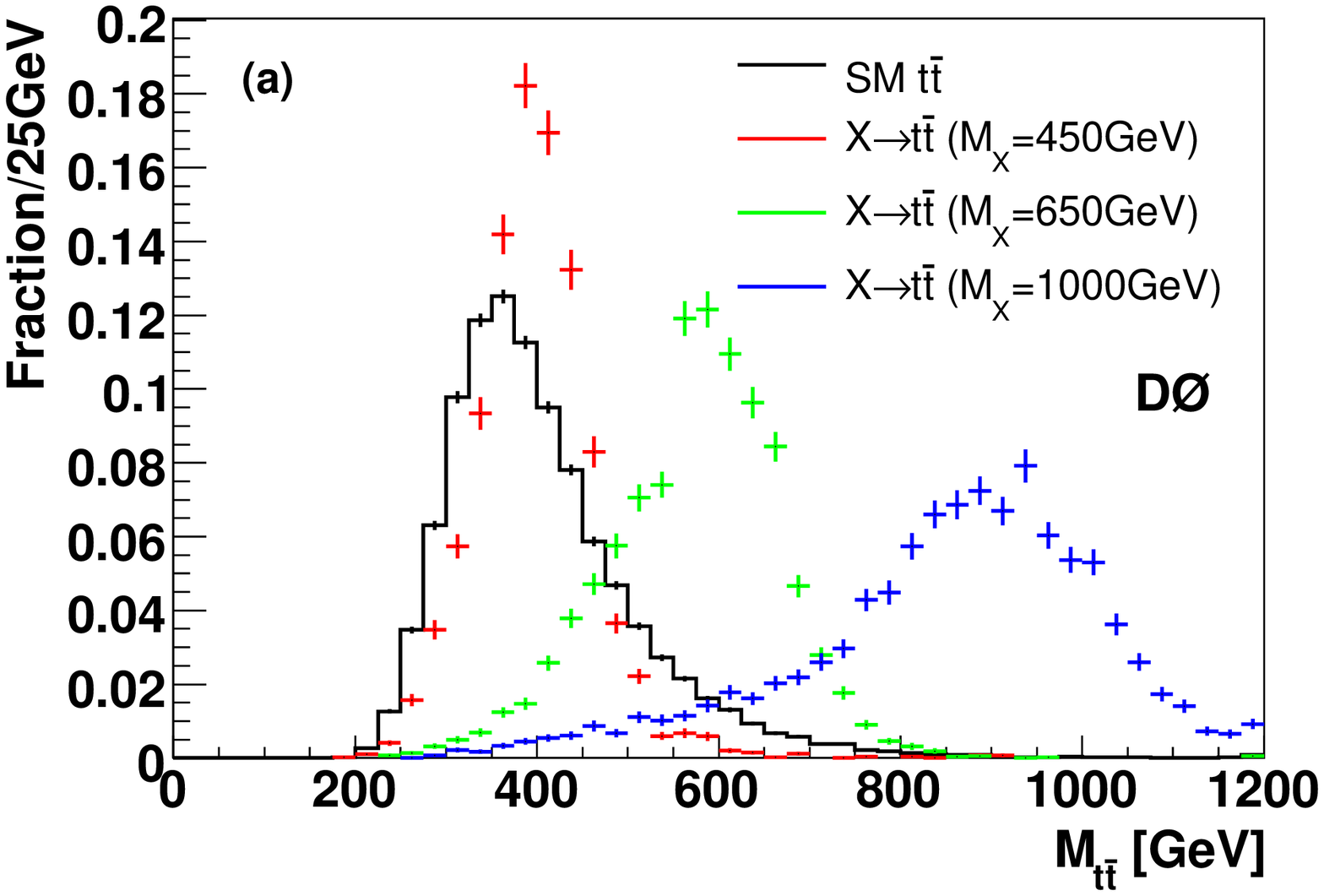}%
\includegraphics[width=0.495\textwidth]{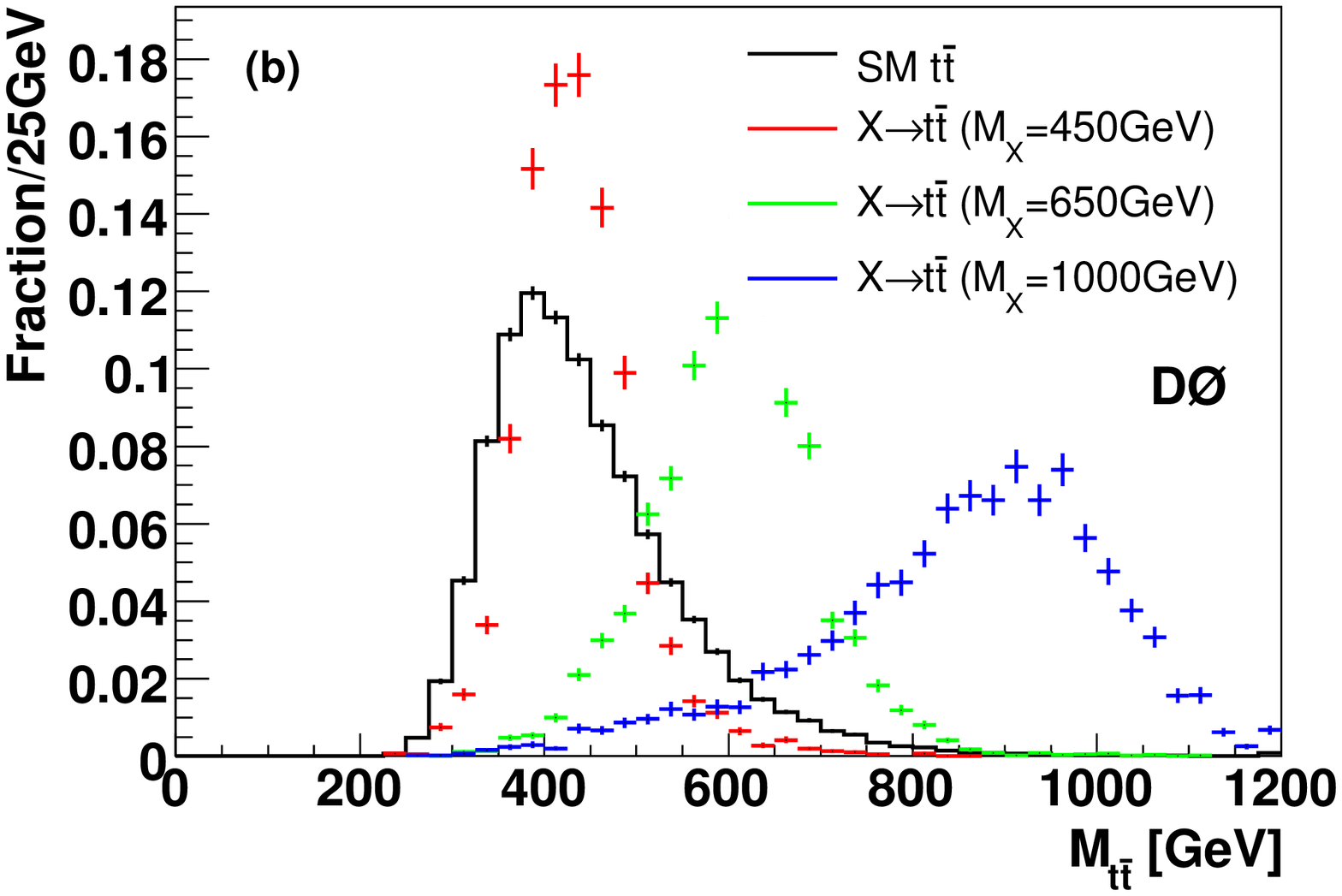}
\caption{\label{fig:zpcomp}Shape comparison of the expected \ttbar invariant mass distributions for 
SM top quark pair production (histogram)
and 
resonant  production from narrow-width resonances of mass 
$M_X=450$, $650$, and $1000\GeV$, 
for (a) $3$ jets events
and (b)~$\ge 4$~jets events.
}
\end{figure*}
The \ttbar invariant mass is reconstructed from the
four-momenta of up to the four highest $p_T$ jets, the lepton momentum, and the
neutrino momentum.
The latter is obtained from the transverse
missing energy and a $W$-mass constraint. The neutrino transverse momentum is identified with the missing
transverse momentum, given by \met and its direction. The neutrino momentum
along the beam direction, $p^\nu_z$, is estimated by solving the equation
$M_W^2=(p^\ell+p^\nu)^2$, where $p^\ell$ ($p^\nu$) is the lepton (neutrino) four momentum.
If there are two solutions, the one with
the smaller $\left|p_z^\nu\right|$ is taken; if no solution exists,  $p_z^\nu$ is
set to zero.
This method  gives better sensitivity for high mass resonances than
a previously applied  constrained kinematic fit
technique~\cite{RunIttresdzero}, 
while only slightly reducing the sensitivity for lower resonance masses.
Moreover,  this direct reconstruction allows the inclusion of data with fewer than
four jets in the case that some jets are
merged, further increasing the sensitivity.
The expected \ttbar invariant mass distributions for three different resonance 
masses are compared to the SM expectation in Fig.~\ref{fig:zpcomp}.

\section{\label{systematics}Systematic Uncertainties}

The systematic uncertainties can
be classified as those affecting only normalization and those affecting 
the shape of any of the signal or background invariant mass distributions.
The systematic uncertainties affecting only the normalization include
the theoretical uncertainty on the SM prediction for $\sigma_{\ttbar}$
($9\%$), 
the uncertainty on the integrated 
luminosity ($6.1\%$)~\cite{Andeen:2007zc},
and the uncertainty on the lepton identification efficiencies.

The systematic uncertainties affecting the shape of the invariant mass
distribution as well as
the normalization are studied in signal and background samples.
These include uncertainties on the jet energy calibration, jet reconstruction efficiency, and $b$-tagging
parameterizations for $b$, $c$ and light jets. The effect due to the top quark mass uncertainty
is computed by changing $m_t$ in the simulation of \ttbar to $165$\GeV\ and $185$\GeV, 
normalized to their corresponding theoretical cross sections.
The effect is scaled to correspond to a top quark mass uncertainty of $\pm 5\GeV$.
The difference in the \ttbar acceptance  due to the top quark mass variation is also
included in the systematic uncertainty.

The fraction of heavy flavor in the \wplus background is measured
in control samples, and a corresponding uncertainty on the \wplus flavor
composition is used. Also the uncertainties 
on the $b$-fragmentation
and the uncertainties of the efficiencies used in the matrix
method are taken into account.

\newcommand{\LEPTID}{Selection efficiencies
}

\begin{table}[b]
\caption{The relative systematic uncertainties on the overall
  normalization of the SM background and for a resonance mass of
  $M_X=650\GeV$, with at least one $b$-tagged jet. The
  uncertainties shown are symmetrized. The actual asymmetric uncertainties and
  the effect of shape-changing systematic errors are used in the limit setting.\label{tab:normsys}}
\begin{tabular}{lrrrr}
\hline
\hline
      &\multicolumn{2}{c}{SM processes}&\multicolumn{2}{c}{Resonance}\\
Source&\multicolumn{2}{c}{(backgrounds)}&\multicolumn{2}{c}{$M_X=650\GeV$}\\
\hline
&\multicolumn{1}{c}{~~$3$ jets~~}&\multicolumn{1}{c}{$\ge 4$ jets}&\multicolumn{1}{c}{~~$3$ jets~~}&\multicolumn{1}{c}{$\ge 4$ jets}\\
\hline
Jet energy calibration                   & $ \pm    1.0\% $ & $ \pm    5.8\% $ &  $ \pm    3.7\% $ & $ \pm    5.5\% $  \\
Jet energy resolution                    & $ \pm    0.2\% $ & $ <    0.1\% $ & $ \pm 1.2\% $ & $ \pm    0.2\% $  \\
Jet identification                       & $ \pm    0.6\% $ & $ \pm    2.0\% $ &  $ \pm    0.6\% $ & $ \pm    1.6\% $  \\
$\sigma_{t\bar t}(m_t\!=\!175\GeV)$       & $ \pm    3.1\% $ & $ \pm    5.9\% $ & $   -~~~ $ & $     -~~~ $  \\ 
Top quark mass                           & $ \pm    5.2\% $ & $ \pm    6.9\% $ & $    -~~~ $ & $     -~~~ $  \\ 
$b$ tagging                              & $ \pm    3.1\% $ & $ \pm    4.9\% $ & $ \pm    3.9\% $ & $ \pm    3.6\% $ \\
$b$ fragmentation                        & $ \pm    0.3\% $ & $ \pm    0.4\% $ &  $ \pm    0.6\% $ & $ \pm    0.6\% $  \\
$W+$jets (heavy flavor)                  & $ \pm    2.5\% $ & $ \pm    0.9\% $ & $    -~~~ $ & $     -~~~ $  \\ 
Multijet  lepton fake rate               & $ \pm    0.3\% $ & $ <0.1\% $ & $     -~~~ $ & $     -~~~ $  \\ 
\LEPTID                                   & $ \pm    3.1\% $ & $ \pm    5.3\% $ &  $ \pm    3.6\% $ & $ \pm    3.6\% $  \\
Luminosity                               & $ \pm    2.6\% $ & $ \pm    4.2\% $& $ \pm    6.1\% $ & $ \pm    6.1\% $  \\ 
\hline
\hline
\end{tabular}
\end{table}

Table~\ref{tab:normsys} gives a summary of the relative systematic uncertainties on the total
SM background normalization for the combined $\ell$+jets channels.
The effect of the different systematic uncertainties on the shape of the \ttbar invariant mass
distribution cannot be inferred from this table, but is included in the analysis.

\begin{figure*}[t]
\begin{center}
\includegraphics[width=0.495\textwidth]{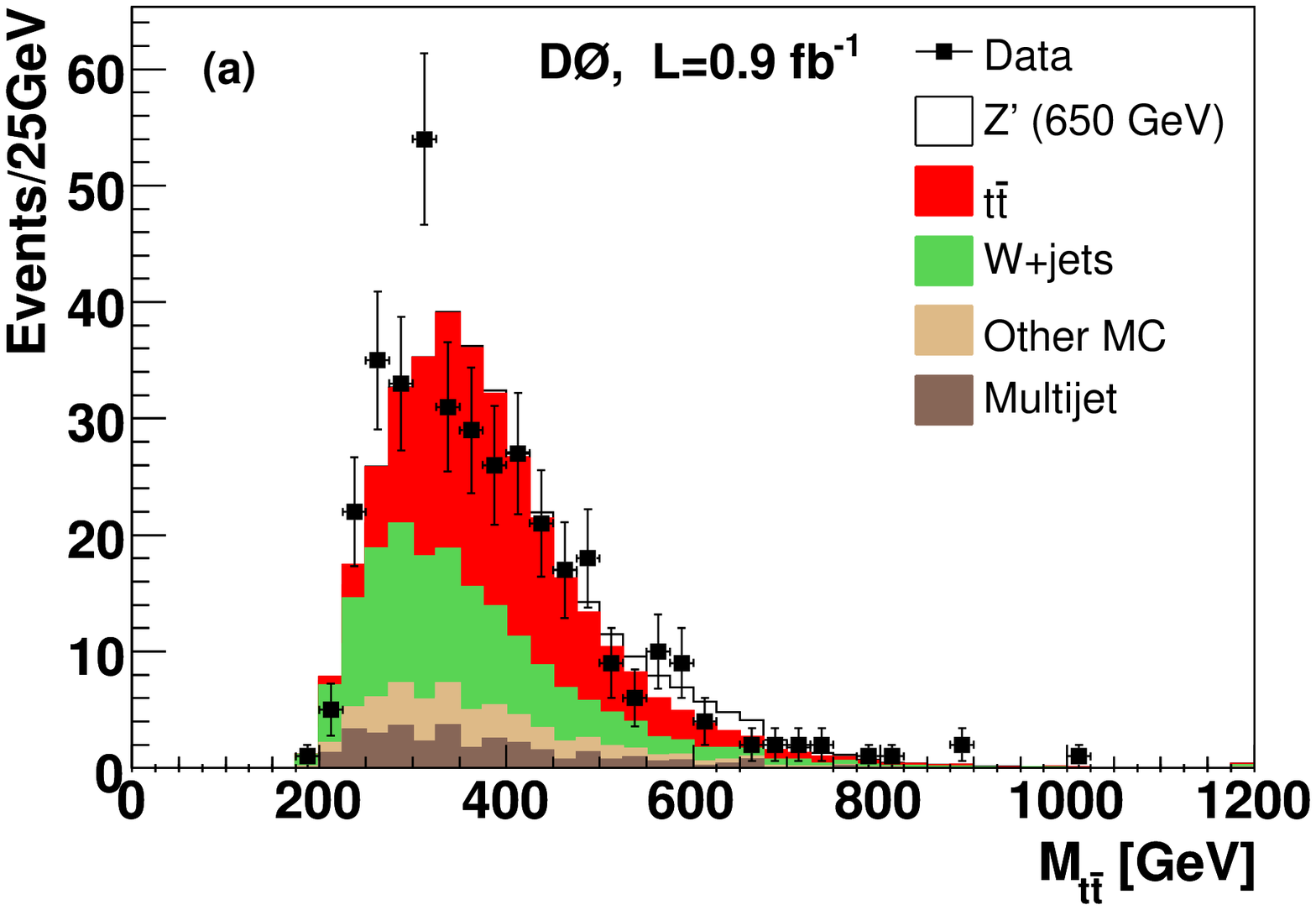}
\includegraphics[width=0.495\textwidth]{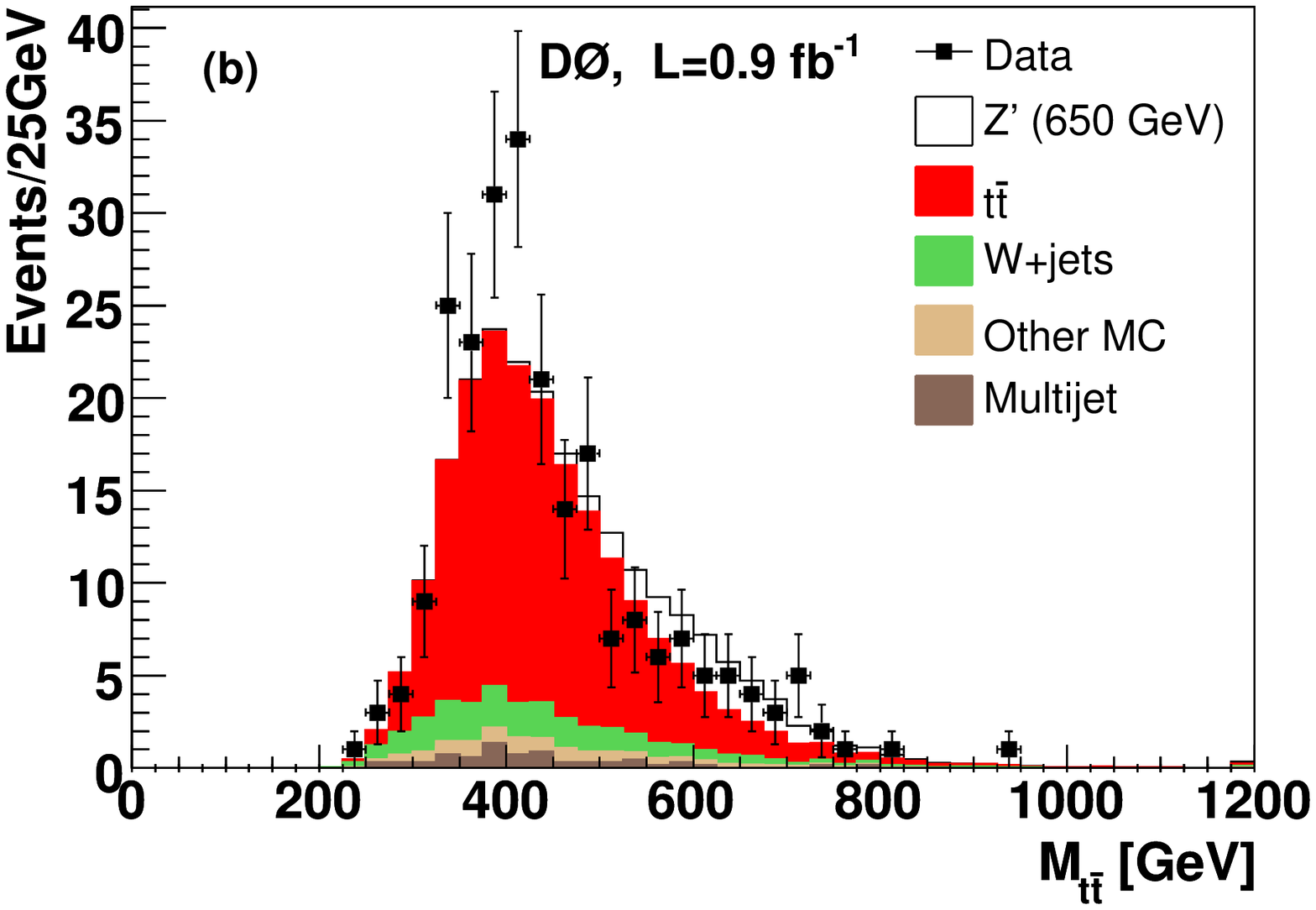}\\[-1em]
\caption{\label{fig:mtt} Expected and observed \ttbar invariant mass
  distribution for the combined (a) $\ell+3$~jets and (b) $\ell+4$ or more jets
  channels, with at least one identified $b$ jet.
Errors shown on the data points are statistical.
Superimposed as white area is the expected signal for a
Topcolor-assisted technicolor $Z'$ boson with $M_{Z'}=650\GeV$.
}
\end{center}
\end{figure*}

\section{\label{result}Result}

After all selection cuts, 319 events remain in the \eplus channel and 288
events in the \muplus channel. The sums of all SM and multijet instrumental
backgrounds are $303\pm22$ and $251\pm19$ events, respectively.
Invariant mass distributions are computed
for events with exactly one $b$ tag and for events with more than one $b$ tag.
Additionally, the distributions are  separated into 3 jets and $\ge 4$ jets samples. 
The measured invariant mass
distributions and corresponding background estimations are shown in  
Fig.~\ref{fig:mtt} for the 3 jets and $\ge 4$ jets samples.

Finding no significant deviation from the SM expectation, we apply
a Bayesian approach to calculate $95\%$ C.L. upper limits on 
{\sigmaB} for hypothesized values of $M_X$ between $350$ and
$1000\GeV$.
A Poisson distribution is assumed for the number of observed events in each bin, and flat
prior probabilities are taken for the signal cross section times
branching fraction. The prior for
the combined signal acceptance and background yields is a multivariate
Gaussian with uncertainties and correlations described by a covariance
matrix~\cite{Bertram:2000br}.

\begin{figure}[b]
\begin{center}
\includegraphics[width=\linewidth]{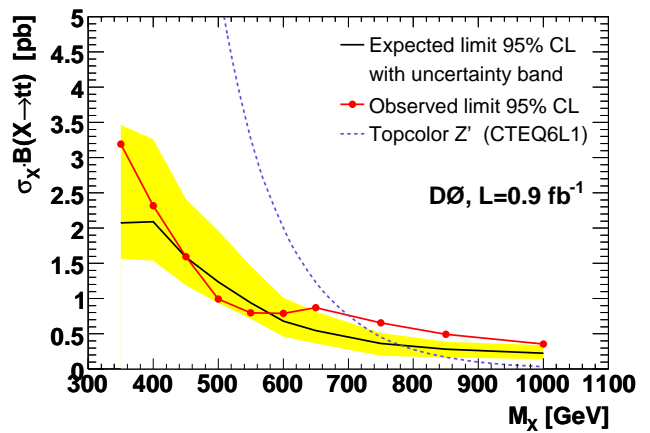}
\caption{\label{fig:limsum}Expected and observed 95$\%$ C.L. upper limits on {\sigmaB} compared with 
the predicted Topcolor-assisted technicolor cross section for a $Z'$ boson with a width of 
$\Gamma_{Z'} = 0.012 M_{Z'}$ as a function of resonance mass $M_{X}$. The shaded band gives the $\pm$1 sigma
uncertainty in the SM expected limit.}
\end{center}
\end{figure}

The expected and observed $95\%$~C.L. upper limits on {\sigmaB} as a function
of $M_X$, after combining the 1 and 2  $b$-tag samples and the 3 and $\ge$ 4 jets samples, are summarized in 
Table~\ref{tab:limit_stat} and displayed in Fig.~\ref{fig:limsum}. This figure also includes the
predicted {\sigmaB} for a leptophobic $Z'$ boson with $\Gamma_{Z'} = 0.012
M_{Z'}$ computed using CTEQ6L1 parton distribution functions. 
The comparison of the observed cross section limits with the $Z$' boson
prediction excludes $M_{Z'} < 700\GeV$ at the $95\%$~C.L.
Due to  a small excess of data over expectation (of no more than
$1.5\sigma$ significance) for invariant masses in
the range between 600 and 700\GeV, the observed limits do not reach the expected
limit for a $Z'$ boson of 780\GeV.

\begin{table}[b]
\begin{center}
\caption{Expected and observed limits for {\sigmaB} at the
    $95\%$~C.L. when combining all channels and
    taking all systematic uncertainties into account.
\label{tab:limit_stat} }
\begin{tabular}{ccc}
\hline
\hline
$M_{X}$ [GeV] & Exp. limit [pb]& Obs. limit [pb]  \\
\hline
 350 &   2.08 &  3.19 \\
 400 &   2.09 &  2.32 \\
 450 &   1.59 &  1.59 \\
 500 &   1.24 &  0.99 \\
 550 &   0.94 &  0.80 \\
 600 &   0.68 &  0.79 \\
 650 &   0.55 &  0.87 \\
 750 &   0.36 &  0.66 \\
 850 &   0.28 &  0.49 \\
1000 &   0.22 &  0.36 \\
\hline
\hline
\end{tabular}
\end{center}
\end{table}

\clearpage

\section{\label{conc}Conclusion}
A search for a narrow-width heavy resonance decaying to \ttbar in the \lplus
final states has been performed using data 
corresponding to an integrated luminosity of about $0.9\ifb$, collected with the \dzero detector
at the Tevatron collider. By analyzing the reconstructed \ttbar invariant mass 
distribution and using a Bayesian method, model independent upper limits on {\sigmaB} have 
been obtained for different hypothesized masses of a narrow-width heavy resonance decaying into \ttbar. 
Within a Topcolor-assisted technicolor model, the existence of a leptophobic $Z'$ boson with 
$M_{Z'} < 700$\GeV\ and width $\Gamma_{Z'} = 0.012 M_{Z'}$ is excluded at
the $95\%$ C.L.

%
We thank the staffs at Fermilab and collaborating institutions, 
and acknowledge support from the 
DOE and NSF (USA);
CEA and CNRS/IN2P3 (France);
FASI, Rosatom and RFBR (Russia);
CNPq, FAPERJ, FAPESP and FUNDUNESP (Brazil);
DAE and DST (India);
Colciencias (Colombia);
CONACyT (Mexico);
KRF and KOSEF (Korea);
CONICET and UBACyT (Argentina);
FOM (The Netherlands);
STFC (United Kingdom);
MSMT and GACR (Czech Republic);
CRC Program, CFI, NSERC and WestGrid Project (Canada);
BMBF and DFG (Germany);
SFI (Ireland);
The Swedish Research Council (Sweden);
CAS and CNSF (China);
and the
Alexander von Humboldt Foundation.
%

\bibliographystyle{utphys_eprintModifiedNILS_DW}
\bibliography{note}

\end{document}